 standard latex file, no figures

==================
\documentstyle[12pt]{article}
\newcommand{\be}{\begin{equation}}
\newcommand{\ee}{\end{equation}}
\newcommand{\bq}{\begin{eqnarray}}
\newcommand{\eq}{\end{eqnarray}}

\begin{document}

\pagestyle{empty}
\begin{flushright}
{\large UBCTP 92-17\\
May 1992}
\end{flushright}
\vspace{0.4cm}
\begin{center}
{\large \bf ABOUT SOME EXACT SOLUTIONS FOR $2+1$  GRAVITY
 COUPLED TO GAUGE FIELDS }\\ \vspace{1 cm}
{\large Ian I. Kogan} \footnote{ On leave of ITEP, Moscow, USSR.},
\footnote{This work is supported in part by the Natural Sciences and
Engineering Research Council of Canada}\\
\vspace{0.6 cm}
{ Department of Physics, University of British Columbia\\
Vancouver, B.C., Canada V6T1Z1}\\
 \end{center}
\vspace{0.4cm} \noindent
\begin{center}
{\bf Abstract}
\end{center}
Some exact  static solutions for  Einstein
 gravity in $2+1$
 dimensions coupled to abelian gauge field are discussed,
 where  the invariant interval is of the
 form: $ds^{2} = N^{2}(r)dt^{2} - dr^{2} - C^{2}(r)d\theta^{2}$.
  Some of these solutions are three-dimensional analogs of the
 Schwarzschild black holes. The  metrics in the
 regions inside and outside the
 horison are connected by the  changing of the Planck mass sign.

\newpage
\pagestyle{plain}
\setcounter{page}{1}
It is known that Einstein gravity in $2+1$ dimensions has no local degrees
 of freedom and outside the matter the space-time is flat. However,
 by coupling gravity to the matter one can get non-trivial space-times.
 For example, by  coupling point particles to gravity one gets
 static  space-time
 with conical singularities \cite{1},\cite{2}. Exact
 non-static solutions in the case of point particles with spin are also
 known \cite{2}. In all these cases the space-time is flat outside the point
 particles.

 In this paper we shall consider some exact non-flat solutions for the
 $2+1$ Einstein gravity coupled to abelian  topologically
 massive gauge field  \cite{3} with action
\bq
S = \int d^{3}x\{\frac{1}{\kappa} \sqrt{-g}R +
 \frac{1}{2}\sqrt{-g}F_{\mu\nu}F^{\mu\nu}
 - m\epsilon^{\mu\nu\lambda}F_{\mu\nu}A_{\lambda} +
 \sqrt{-g}J^{\mu}A_{\mu}\}
\eq
 where $J^{\mu}$ is the covariantly conserved current: $D_{\mu}J^{\mu}=0$.
  The coupled Einstein-Maxwell equations are
\bq
R_{\mu\nu} - \frac{1}{2} g_{\mu\nu} R = \kappa T_{\mu\nu} \nonumber \\
\partial_{\nu}(\sqrt{-g}F^{\nu\sigma}) + m\epsilon^{\sigma\mu\nu}F_{\mu\nu}
 = \sqrt{-g}J^{\sigma}
\eq
where the stress-energy tensor $T_{\mu\nu}$ does not depend on the
 gauge Chern-Simons term and equals to
\bq
T_{\mu\nu} = - F_{\mu\rho}F_{\nu}^{\rho} + \frac{1}{4}g_{\mu\nu}
F_{\lambda\sigma}F^{\lambda\sigma}
\eq

We shall looking for solutions which depend only on radial coordinate $r$ only,
then the metric can be represented as
\bq
ds^{2}  = N^{2}(r)dt^{2} - dr^{2} - C^{2}(r)d\theta^{2}
\label{metric}
\eq
 and only non-zero $F_{\mu\nu}$ components are  electric $F_{0r} = E(r)$
 and magnetic $F_{r\theta} = H(r)$ fields. It is easy to see that from
 $D_{\mu}J^{\mu} = 0$ one gets $J^{r} = 0$.
After simple calculations one gets ($X' = dX/dr$):
\bq
R_{00} & =  & NN'' + N N' \frac{C'}{C}=\kappa \frac{N^{2}}{C^{2}}H^{2}
 \nonumber \\
R_{rr} & =  & -\frac{N''}{N} - \frac{C''}{C}=0 \\
R_{\theta\theta}  & = & - CC'' - CC'\frac{N'}{N}=
 \kappa\frac{C^{2}}{N^{2}}E^{2}
 \nonumber
\label{1}
\eq
and
\bq
\frac{d}{dr}(\frac{C}{N} E) + mH = CNJ^{0} \nonumber \\
\frac{d}{dr}(\frac{N}{C} H) + mE = CNJ^{\theta}
\label{2}
\eq
It is easy to see from (\ref{1}), (\ref{2})
 that Eistein-Maxwell equations are symmetric under
 the transformation
\bq
N \leftrightarrow C, \;\; E \leftrightarrow H, \;\;
J^{\theta} \leftrightarrow J^{0},\;\; \kappa \rightarrow
 -\kappa
\eq
which is easy to understand because of the formal
 symmetry between $\theta$ and $it$ in (\ref{metric}).

However, one can  put $J^{\theta} =0$ and get two solutions  with
 $E \sim N/C,\; H=0$ and $H \sim C/N,\; E=0$ considering point-like
 charge in pure Maxwell theory, i.e. Chern-Simons mass term is zero,
 $m=0$ or uniform  charge distribution in the topologically massive
 gauge theory with non-zero Chern-Simons term
 $m \neq 0$. Then we see that the abovementioned
 duality can be realised not as duality betwen time and angular components
 of the current in the same theory, but as a duality between
 point-like charge distribution in pure Maxwell theory and uniform
 charge distribution in topologically massive gauge theory.

 Let us consider the  first case $m=0$. It is easy to see from (\ref{2})
 that outside the source $J^{0}$
 one gets
\bq  E = \frac{N}{C} Q , \;\; H=0
\label{E} \eq
 with constant $Q$  . For point-like
 charge $J^{0} = Q(CN)^{-1}\delta(r)$ and $Q$ is a charge. However one
 can get (\ref{E}) as a solution without sources if the resulting
 space-time manifold will be free of singularities.

In the second case $m \neq 0$ one can easily gets the
 solution
\bq
H = \frac{C}{N} Q, \;\; E =0
\eq
where now $J^{0} = mQ/N^{2}$, i.e. $J_{0} = N^{2}J^{0} = mQ = const$
 which corresponds to the  uniform distribution of the  charge.

Now let us consider the solutions of (\ref{1}) in electric case.
Einstein equations take the form:
\bq
 NN'' + N N' \frac{C'}{C} &= & 0 \nonumber \\
\frac{N''}{N} + \frac{C''}{C} &= & 0 \\
 - CC'' - CC'\frac{N'}{N} & = &
 \kappa Q^{2} \nonumber
\label{1E}
\eq

It is easy to see that from (10) one gets
\bq
2CC''& =  & -\kappa Q^{2} \nonumber \\
 \frac{N'}{N} & = & \frac{C''}{C'}
\eq
{}From the second equation we get $N \sim C'$ and the first equation has
 the first integral
\bq
(C')^{2} + \kappa Q^{2} ln\frac{C}{C_{0}} = 0
\eq
 where $C_{0}$ is the integration constant which we shall put equal to $1$
 after  the resclaing of $\theta$.   Thus the
 integration of the Einstein equations is reduced to the integration
 of the particle's movement in the logarithmic potential. In the
 case of positive Planck constant $\kappa > 0$ the movement is restricted
 in the region $(0,1)$ (Fig.1), in the case of the negative
 $\kappa < 0$ \footnote{let us note that it is this case which corresponds
 to the low-energy limit of the topologically massive gravity \cite{3}} $C(r)$
 takes the value in the interval $(1, \infty)$ (Fig.2).

Let us consider the  case $\kappa > 0$. The first integral (12) takes the
  form
\bq
(\frac{dC}{d\tau})^{2} + ln C = 0, \;\; \tau = Q\sqrt{\kappa} r
\eq
 It is convenient to substitute $C = \exp(-F^{2})$, then one
 gets
\bq
\tau(F) = 2\int_{0}^{F} dF e^{-F^{2}}
\eq
The qualitative behavior $C(r)$ is drawn on Fig.3 - we get two singular
 points at at  $r=  r_{\pm} = \pm\tau(\infty)/Q\sqrt{\kappa} =
 \pm(1/Q)\sqrt{\pi/
\kappa}$.
It is easy
 to get asymptotics  near $r=0$ and $r_{\pm}$ :
\bq
C(r)  &=& exp[-\frac{Q^{2}\kappa}{4} r^{2} + O(r^{4})], \;\;
 r \rightarrow 0; \nonumber \\
C(r)  &=&  Q\sqrt{\kappa}\epsilon (-\ln Q\sqrt{\kappa}\epsilon)^{1/2}, \;\;
\epsilon =
 r_{+}-r,\; r-r_{-}.
 \rightarrow 0;
\eq

 One can consider $C$ itself as a radius, then the angular part has normal
 form $C^{2} d\theta^{2}$, but the radial part becomes $dr^{2} \sim dC^{2}/
 \ln(1/C)$.
 Let us note that  $r= 0$ is the extremal point for $C(r)$ and thus
 $N \sim C' = 0$ at this point, i.e. one can thinks that there is a horison.
 Near the singular points $N$ has logarithmic singularity .
 Thus, in the positive $\kappa$ case we get the closed universe - contrary
 to the $3+1$ case where the energy density is small at large distances
 (like $1/r^{4}$ in the Reissner-Nordstrem black hole) and space-time
 becomes asymptotically flat in the $2+1$ case  $E^{2} \sim 1/r^{2}$ in
 flat space and the total energy $\int rdr E^{2}$ diverges at large $r$.
 This is the simple explanation why at some finite $r$ we get the second
 singularity - elecric field energy is large enough to close the space.

 Let us consider now case $\kappa <0$. In this case one gets
\bq
(\frac{dC}{d\tau})^{2} - ln C = 0, \;\; \tau = Q\sqrt{|\kappa|} r
\eq
Substituting $C =  \exp(F^{2})$,  one
 gets
\bq
\tau(F) = 2\int_{0}^{F} dF e^{F^{2}}
\eq
The qualitative behavior $C(r)$ is drawn on Fig.4. Now we get the regular
 solution without singularity, because $C$ never reach zero. It is easy
 to get asymptotics  near $r=0$ and $\pm\infty$ :
\bq
C(r)  &=& exp[\frac{Q^{2}|\kappa|}{4} r^{2} + O(r^{4})], \;\;
 r \rightarrow 0; \nonumber \\
C(r)  &=&  Q\sqrt{|\kappa|} |r| (\ln Q\sqrt{|\kappa|} |r|)^{1/2}, \;\;
 r \rightarrow  \pm\infty;
\eq
We see that the space is not asymptotically flat - the same problem with
 gauge field energy  which is not enough small at large distances,
 but now due to negative $\kappa$ it makes space more "open" that the flat
 space. Let us note that at $N \sim C' \sim r$ and it means that there is
 a horison at $r=0$. The negative $r$ region is unphusical - in some sense
 the correct radial coordinate is $r^{2}$, not $r$ and the metric we
 got describes the region $r>0$ outside the horison.

It easy to see that in both cases
 the space-time at $r>0$ is not geodesically complete, which is the
 necessary condition for $r=0$ to be a horison.  To demonstrate it
 let us consider the equation for geodesic line $r = r(s), t= t(s), \theta
 = const$:
 \bq
\frac{d^{2}x^{\mu}}{ds^{2}} + \Gamma^{\mu}_{\nu\rho}\frac{dx^{\nu}}{ds}
\frac{dx^{\rho}}{ds} =0
\eq
For our metric the only nonzero $\Gamma$ with $r$ and $t$ indices are
\bq
\Gamma^{t}_{tr} = \frac{N'}{N},\;\;\; \Gamma^{r}_{tt} = NN'
\eq
and we get
\bq
\frac{d^{2}r}{ds^{2}} + NN'(\frac{dt}{ds})^{2} &= &0 \nonumber \\
\frac{d^{2}t}{ds^{2}} + \frac{N'}{N}\frac{dt}{ds}\frac{dr}{ds} &= &0
\eq
 Using the fact that $N'(dr/ds) = dN/ds$ one gets from
 the second equation $N(dt/ds) = const$, then  the
 first equation can be written as
\bq
\frac{d^{2}r}{ds^{2}} + \frac{d \ln N}{dr} = 0 \nonumber \\
 \frac{1}{2}(\frac{dr}{ds})^{2} + \ln N = const
\eq
and we get the equation of motion in potential $\ln N$.
 One can fix the normalization of $N$ as $N = C'/M$, where
 $M=Q\sqrt{|\kappa|}$ (any other normalization can be obtained from
 this one by the resclaing of time $t$), then near the
 horison $r=0$ one gets $N = Mr/2$ (in the case of positive $\kappa$ when
 $C' <0$ near horison we simply change the sign of $N$ to make it positive).
 Thus we again have the
 problem of the logarithmic potential as in the case of positive
 $\kappa$ (see Fig.1), but now not for the metric itself, but for
the radial
 coordinate of the free moving particle. We know that the particle
 can reach the origin in
 the finite proper time $s$ and this means that $r=0$ is indeed horison
 - the free falling observer can reach it in a finite  proper time,
 but from the point of view of the external observer at some fixed $r>0$
 this process takes infinite time. It is easy to write the exact expression
 for the geodesic line $dr/dt = const N \sqrt{-2\ln (N/N_{0})}$,
 where $N_{0}$ is some constant. Using  $N = Mr/2$ near the horison
 we get $r = c_{1}exp(-c_{2}t^{2})$ and $r\rightarrow 0$ only at
 $t \rightarrow \infty$.

Before discussing  the metric inside the horison let us consider
 the Eucledian
version of this metric
\bq
ds^{2} = N^{2}dt^{2} + dr^{2} + C^{2} dr^{2}
\eq
Near the horison $N = Mr/2$ and we see that imaginary time $t$
becomes angular
 variable $ t= (2/M)\phi$ \footnote{let us note that because $C
\neq 0$ for negative $\kappa$ our original angular
 variable $\theta$ may be indeed noncompact variable - one get
 then periodicity
 in $\theta$ by factorising universal covering space by arbitrary
 period}
with period $4\pi/M$  in a complete analogy with the ordinary
 four-dimensional Eucledian Schwarzschild metric \cite{4}.
Then horison
 becomes the regular point and we obtain the regular 3-dimensional manifold
 without singularities in the case $\kappa <0$. For positive $\kappa$
 there is naked singularity at $r=r_{+}$.

It is convenient to consider new  (Kruskal) variables
\bq
U = e^{\rho + i\phi}, \;\; \bar{U} = e^{\rho - i\phi}
\eq
where new radial coordinate $\rho$ is defined as
\bq
\frac{dr}{d\rho} = \frac{2}{M} N = \frac{2}{M^{2}}\frac{dC}{dr}
\label{rhor}
\eq
 Then the metric takes the form ($C' = dC/dr$)
\bq
ds^{2} = \frac{4C'^{2}}{M^{4}U\bar{U}} dUd\bar{U} + C^{2} d\theta^{2}
\eq
Using (13), (16) it easy to see that
 $\frac{4C'^{2}}{M^{4}} = 4 F^{2}/M^{2}$ where $F$ was defined
  as $C = exp(-F^{2})$ for $\kappa>0$ and $C=exp(F^{2})$ for
$\kappa <0$
Unfortunately we can not obtain the explicit function $F = F(U\bar{U})$,
 however one can easily find the asymptotics at small and large $|U|$.

Let us first consider negative $\kappa$, where $0<U\bar{U}<\infty$.
Using (\ref{rhor}) one can get $C'' = (M^{2}/2)(dC/d\rho)$ and from (16)
we get
\bq
\frac{exp(F^{2})}{F}\frac{dF}{d\rho} = 1
\label{+F}
\eq
 After some calculations one gets:
\bq
4F^{2}/M^{2} =  \frac{4}{M^{2}} e^{2\rho} =
\frac{4}{M^{2}}U\bar{U} ,\;\;
  U\bar{U}<<1 \nonumber \\
4F^{2}/M^{2} = \frac{4}{M^{2}}(\ln\ln U\bar{U}+ \ln\ln\ln U\bar{U}) , \;\;
, \;\; U\bar{U}>>1
\eq
The Eucledian metric
\bq
ds^{2} = \frac{4F^{2}(U\bar{U})}{M^{2}U\bar{U}} dUd\bar{U} + e^{2F^{2}}
 d\theta^{2}
\eq
is regular at $U\bar{U} = 0$

Now let us consider the positive $\kappa$. Because now $C' <0$  we take
(25) with negative sign and then  can rewrite  (13) as
\bq
\frac{exp(-F^{2})}{F}\frac{dF}{d\rho} = 1
\label{-F}
\eq
in an anology with  equation (\ref{+F}).
At small $|U|$ one gets the same $F$ as in the case $\kappa <0$, but now
 the is singularity at some finite $U\bar{U} = e^{A}$, where the numerical
 value of $A$ is
\bq
A = 2 \int_{1}^{\infty} \frac{e^{-x^{2}}}{x} dx
\eq
The Eucledian metric for positive $\kappa$ is
\bq
ds^{2} = \frac{4F^{2}(U\bar{U})}{M^{2}U\bar{U}} dUd\bar{U} + e^{-2F^{2}}
 d\theta^{2}
\eq
and is singular at $U\bar{U} = e^{A} $

Now let us analytically continue back to Minkowski space-time $t
\rightarrow it, \; \phi \rightarrow i\phi$. Instead of complex conjugate
 $U$ and $\bar{U}$ one gets  real light-cone coordinates $U$ and $V$:
\bq
U = e^{\rho -\phi}, \;\; V = e^{\rho + \phi}\;\; U\bar{U}
\rightarrow UV
\eq
Now, contrary to Eucledian space, $UV$ may be negative as well as positive
 and negative $UV$ describes the region inside the horison - in complete
 analogy with Schwarzschild black holes in Kruskal coordinates.
To get the metric inside the horison, i.e. at negative $UV$ let us consider
equation connecting $F$ and $UV$ which can be
 obtained from  (see (\ref{+F}), (\ref{-F})) by substituting $U\bar{U}$ to
 $UV$ in the region where $UV > 0$:
\bq
\frac{exp(\mp F^{2})}{F^{2}} dF^{2} = \frac{ 1}{UV}d (UV)
\eq
where $-$ corresponds to positive $\kappa$ and $+$ for negative one.
This equation can be easily continuued to the negative $UV$  after
 changing the sign of $F^{2} \rightarrow - F^{2}$. But this procedure
simply exchange our two solutions - if one starts from some $\kappa$
(positive or negative) and gets  metric outside the horison $UV >0$
(we write $-ds^{2}$ to restore our initial signature (+,-,-))
\bq
-ds^{2} = \frac{4F^{2}(UV)}{M^{2}UV} dUdV + e^{ \pm F^{2}}
 d\theta^{2}
 \eq
the same metric describes the space-time inside the horison
 $UV <0$, but with
$F$ corresponding now to opposite $\kappa$ (negative or positive).

 Thus Minkowski solution describes the space-time with the horisons
 at $UV = 0$ and singularities at positive (negative)
  $UV = e^{A}$ in the case of
 positive (negative) Planck mass.  The Penrose diagram is as in
 the case of usual black holes (see Fig. 5, where $\kappa < 0$) and the
 only difference from the Schwarzschild case is that space-time is not
 asymptotically flat. Let us note that  the Planck mass sign defines
 the orientation of the Penrose diagram and $\kappa \rightarrow
 -\kappa$ corresponds to the  $\pi/2$ rotation of the Penrose diagram.

 Finally let us discuss the magnetic case - as we know it corresponds to
 the exchange between $t$ and $\theta$ ( $C \leftrightarrow N$)
 and changing the sign for $\kappa$.
Thus,  for positive $\kappa$  with $E=0, H \neq 0$ one gets $C \sim N'$
 and  $N$ approaches some constant at $r =0$, when $C \sim r$. This means
 that $r=0$ is the regular point and $\theta$ is the usual angle. The space
 time has no horisons and is  open, but not asymptotically flat. In the
 case of negative $\kappa$ one gets again the regular point at zero, but
 at some finite $r$ $C$ is singular - our space time has a singularity at
 finite  radius. More detailed analysis of this situation can be done
 in a complete analogy with the electric case - let us note the
Eucledian versions  are the same as in electric case after obvious
 exchange $t \leftrightarrow \theta$, but Minkowski space-times are different -
 there are no horisons, instead we have regular point $r=0$ and $\theta$
 must be correct angular variable, contrary to electric case, where
 $\theta$ was not restricted to be compact.

 In conclusion we would like to discuss some open problems. It will
 be interesting to understand if it is possible to obtain asymptotically
 flat solutions, using for example, massive gauge fields. The second
 important problem is to find the solution with
 point-like charge in the topologically massive gauge theory -
in this case one immediately gets $dtd\theta$ component for metric tensor
 and it is not evident if the full  Einstein-Maxwell system of
 equations can be integrated explicitly.  And finally it will be
 exteremely interesting to look for these solutions in the topologically
 massive gravity - it will definitely change the behaviour at small
 distances and it is unclear if  a horison at $r=0$ will survive. In this
 case  the $dtd\theta$ componet of the metric tensor will also appear.

{\bf Acknowledgments}

 I am grateful to G.Semenoff and  N.Weiss
 for interesting  discussions and  hospitality at
  the University of British
Columbia. It is a pleasure to acknowledge helpful
 discussions with  S.Carlip and  W.Unruh.

\end{document}